\documentclass
[12pt]{article}
\usepackage[dvips]{color}
\usepackage{epsfig}
\usepackage{amsmath}
\usepackage{graphicx}
\begin{document}
\title {Jet Quenching Parameter with Hyperscaling Violation }
\author{J. Sadeghi $^{a}$\thanks{Email:
j.sadeghi@umz.ac.ir}\hspace{1mm} and S. Heshmatian
$^{a}$\thanks{Email: s.heshmatian@umz.ac.ir}\\
$^a$ {\small {\em  Sciences Faculty, Department of Physics, Mazandaran University,}}\\
{\small {\em P. O. Box 47415-416, Babolsar, Iran}}\\
{\small {\em  Institute for Studies in Theoretical Physics and Mathematics (IPM)}}\\}
\maketitle
\begin{abstract}
In this paper we study the behavior of jet quenching parameter in the background metric with hyperscaling violation at finite temperature.The background metric is covariant under a generalized Lifshitz scaling symmetry with the dynamical exponent $z$ and hyperscaling exponent $\theta$ . We evaluate the jet quenching parameter for certain range of these parameters consistent with the Gubser bound conditions in terms of $T$, $z$ and $\theta$. We compare our results with those from conformal case and experimental data. Then we add a constant electric field to this background and find its effect on the jet quenching parameter.\\
\end{abstract}
\section{Introduction}
Holography is a powerful tool to map a $D$ dimensional strongly coupled field theory at large $N$ limit to a $D+1$ dimensional gravitational theory at weak coupling [1-2].
In recent years, the gauge/gravity duality has been used to study the QCD and hadron physics.
The dynamics of moving quark and the motion of a quark-antiquark pair in a strongly coupled plasma have been studied in the context of gauge/gravity duality [3-20]. Also, the computations of QCD parameters demonstrate the efficiency of this duality [16-20].\\
Generalizations of metrics dual to field theories have been proposed because of the extensive applications of this duality. One of such generalizations is to use metrics dual to field theories which are not scale invariant while they are conformal to Lifshitz spacetimes [39-43]. These backgrounds have a dynamical Lifshitz parameter $z$ and a hyperscaling violation exponent $\theta$. Such metrics have been used to describe condense matter systems and string theory solutions [44-58]. Lorentz symmetry represents a foundation of both general relativity and the standard model, therefore Lorentz invariance violation may leads to new physics.\\
Due to the quark confinement, as $q\bar{q}$ pair are moving apart, they will hadronized by creating more pairs and come out as the jets.
Then some of the jets are suppressed by the medium and lose energy which is so called jet quenching phenomenon. The jet quenching parameter is the probability of the jet quenching and is related to momentum fluctuation $\hat{q}\propto\langle\vec{p}_{\bot}^{^{2}}\rangle $ [32-38]. In the dual description, this parameter is related to the coefficient in the exponent of an adjoint Wilson loop [21-31].\\
In this paper, we use the gauge/gravity duality to study the behavior jet quenching parameter in the background metric with hyperscaling violation at finite temperature with and without a constant B-field.
This paper is organized as follows. In section 2, we briefly review the background of ref.[57].  The metric is covariant under a generalized Lifshitz scaling symmetry and has a generic Lorentz violating form. In section 3, we use the holographic description of strongly coupled QFT to illustrate the behavior of the jet quenching parameter at finite temperature. We determine the allowed regions for $z$ and $\theta$ and evaluate the jet quenching parameter numerically in terms of the background Hawking temperature and $z,\theta$ parameters. In section 4, we added a constant B-field (electric filed) to the background metric and estimate the jet quenching parameter dependence on the Hawking temperature and also the electric field.

\section{Review of the background}
In this section, we briefly review the background introduced in ref.[57]. The authors have considered a background metric where Lorentz invariance is broken. It was argued that although charge densities induce
a trivial (gapped) behavior at low energy/temperature, there are special cases where
there are non-trivial IR fixed points (quantum critical points) where the theory is
scale invariant and have a generic Lorentz violating form.
These metrics, in general, have the following form,
\begin{equation}
ds^2=u^{\theta}\left[-{dt^2\over u^{2z}}+{b_0~du^2+dx^idx^i\over u^2}\right]\\,
\end{equation}
which is covariant under a generalized Lifshitz scaling symmetry,
\begin{equation}
 t\rightarrow \lambda^{z}t\,\,\,,\,\,\, u\rightarrow \lambda u\,\,\,,\,\,\, x^i\rightarrow\lambda x^i\,\,\,,\,\,\, ds^2\rightarrow \lambda^{-\theta} ds^2\\.
\end{equation}
The exponent $z$  is the Lifshtz parameter and the exponent $\theta$ is the hyperscaling violation exponent which is responsible for
the non-standard scaling of physical quantities and controls the transformation of the metric.
The scalar curvature correspond to these geometries is given by the following equation,
\begin{equation}
R=-{3\theta^2-4(z+3)\theta+2(z^2+3z+6)\over b_0}~u^{-\theta}\,.
\end{equation}
The geometries are flat for $\theta=2$ and $z=0,1$. The $(\theta=0,z=1)$ geometry is in Ridler coordinates and Ricci flat when $\theta=4$ and $z=3$.
These special solutions violate the conditions of ref.[57].
For $\theta=0$, the scalar curvature is constant  (pure Lifshitz case, ref.[43]).\\
Using the following radial redefinition,
\begin{equation}
u=(2-z)~r^{1\over 2-z}\\,
\end{equation}
and rescaling of $t,x^i$, the following metric is obtained,
\begin{equation}
ds^2\sim r^{\theta-2\over 2-z}\left[-f(r)dt^2+{dr^2\over f(r)}+dx^idx^i\right]\,\,\,,\,\,\, f(r)=f_0~r^{2{1-z\over 2-z}}\\.
\end{equation}
The energy scale is given by the scaling of the $g_{tt}$ component of the metric.
So, at the presence of hyperscaling violations, one can obtain,
\begin{equation}
E\simeq u^{{\theta-2z}\over 2}\simeq {r^{\theta-2z\over 2(2-z)}}\\.
\end{equation}
For the generalized scaling solutions of eq.(5) the Gubser conditions become,
\begin{equation}
{2z+3(2-\theta)\over 2(z-1)-\theta}>0\,\,\,\,,\,\,\,\, {z-1\over 2(z-1)-\theta}>0\,\,\,\,,\,\,\,\, {2(z-1)+3(2-\theta)\over 2(z-1)-\theta}>0\\,
\end{equation}
and the thermodynamic stability implies that,
\begin{equation}
{z\over  2(z-1)-\theta}>0\\.
\end{equation}
Equations (7) and (8) lead to $z(z-1)>0$.
Authors in ref.[57] have also considered several cases for the generalized Lifshitz geometries as,
\begin{itemize}

\item{1a)} {\bf ${\theta-2\over z-2}> 0$} and  ${z-1\over z-2}>0$.
\item{1b)} {\bf ${\theta-2\over z-2}> 0$} and  ${z-1\over z-2}<0$.
\item{2a)} {\bf ${\theta-2\over z-2}< 0$} and  ${z-1\over z-2}>0$.
\item{2b)} {\bf ${\theta-2\over z-2}< 0$} and  ${z-1\over z-2}<0$.
\end{itemize}
In two first cases, the boundary is at $r=0$ and in two later cases, the boundary is at $r=\infty$. For the last case, there are no acceptable geometries surviving the Gubser bounds in eq.(7).\\
The generalization of metrics in eq.(5) to include finite temperature is,
\begin{equation}
ds^2\sim \left({r\over \ell}\right)^{-\alpha}\left[-f(r)dt^2+{dr^2\over f(r)}+dx^idx^i\right],\ \nonumber
\end{equation}
\begin{equation}
f(r)= f_0 \left({r\over \ell}\right)^{2\beta}h\,\,\,\,\,,\,\,\,\,\,
h=1-\left({r\over r_0}\right)^{\gamma}\\,
\end{equation}
where $\alpha = \frac{\theta-2}{z-2} $, $\beta = \frac{z-1}{z-2}$ and $\gamma = \frac{z+{3\over 2}(2-\theta)}{2-z}$.
\section{Jet quenching parameter}
In this section we analyze the behavior of jet quenching parameter for the background metric of equation (9).
In the holographic description, the jet quenching parameter is computed using the Wilson loop joining two light-like lines by the following equation,
\begin{equation}
\langle{\cal W}^A(C)\rangle\simeq
\exp\left(-\frac{1}{4\sqrt{2}}\hat{q}L^-L^2\right)\,,
\end{equation}
where $C$ is the null-like rectangular Wilson loop formed by a dipole with heavy $q\bar{q}$ with small separation length $L$ and large separation length $L^-$ along the light-cone. Using the relations $\langle {\cal W}^F(C)\rangle^2
\simeq \langle{\cal W}^A(C)\rangle$ and $\langle{\cal W}^F(C)\rangle \simeq  e^{-S_I}$, the jet quenching parameter in given by,
\begin{equation}
\hat{q} \equiv 8\sqrt{2}\frac{S_{I}}{L^-L^2}\,,
\end{equation}
where  $S_{I} = S - S_{0}$. Here $S$ is the Nambu-Goto action of the fundamental string, $S_0$ is the
self energy from the mass of two quarks and  $S_{I}$ is the regularized string worldsheet action .\\
To evaluate the jet quenching parameter we start with the background black hole solution, equation (9), and use the following light-cone coordinates,
\begin{equation}
 x^{\pm} = \frac{t\pm x^1}{\sqrt{2}}\,,
\end{equation}
to rewrite the black hole metric as,
\begin{equation}
ds^2=b^{2}(r)[-(1+f(r))dx^+dx^-+\frac{1-f(r)}{2}[(dx^+)^2+(dx^-)^2]
+(dx^2)^2+(dx^3)^2+f^{-1}(r)dr^2]\,,
\end{equation}
where $b^{2}(r) = \left({r/\ell}\right)^{-\alpha}$. We choose the static gauge  $\tau = x^- (0 \leq x^- \leq L^-)$, $\sigma
= x^2 (-\frac{L}{2} \leq x_2\leq \frac{L}{2})$ and consider the string with endpoints to be located
at $\sigma = \pm \frac{L}{2}$. In the limit with $L^- \gg L$, the effect of $x^-$ dependence of the worldsheet
can be neglected and the string profile is completely specified by $r = r(\sigma)$.\\
The induced metric of the fundamental string can be calculated as,
\begin{equation}
g_{\alpha\beta} = b^2(r) \left(\begin{array}{cc}-\frac{f(r)-1}{2}& 0\\ 0 & 1+\frac{r^{'2}}{f(r)}\end{array}\right)\,.
\end{equation}
Plugging the above equation into the Nambu-Gotto action, we obtain,
\begin{equation}
S = - \frac{1}{2 \pi \alpha'} \int d\tau d\sigma \sqrt{-{\rm
det}g_{\alpha \beta}} =  \frac{L^-}{\sqrt{2} \pi \alpha^{'}} \int^{\frac{L}{2}}_{0}  dy
   b^{2}(r)\sqrt{\left(f(r)-1\right)\left(1+\frac{r^{'2}}{f(r)}\right)}\,.
\end{equation}
Since the Lagrangian density does not explicitly depend
on $y$, the corresponding Hamiltonian is conserved and we can write,
\begin{equation}
 \frac{\partial {\cal L}}{\partial r^{'}}r^{'} - {\cal L} = E \\,
\end{equation}
where $E$ is the constant energy of motion and ${\cal L}$ is the integrand of equation (15).
Then we obtain the equation of motion for $r$ as,
\begin{equation}
{r^{'}} = \sqrt{f(r)\left[\frac{b^4(r)(f(r)-1)}{2E^2}-1\right]}\,.
\end{equation}
Due to the fact that for the black hole solution we have $f(r) = 0$ at the horizon and as we are interested in the small $E$ case,
the factor under square root is always positive near the boundary and negative near the black hole horizon.
Therefore, the turning point $r_{min}$ ($>r_h$) is determined by the solution of the following equation,
\begin{equation}
f(r_{min})-1=0\,.
\end{equation}
Substituting eq.(17) in the Nambu-Gotto action of eq.(15), the action becomes,
\begin{equation}
S =  \frac{L^-}{\sqrt{2} \pi \alpha^{'}} \int^{r_b}_{r_{min}}  dr
 b^{2}(r)\sqrt{\frac{f(r)-1}{f(r)}}\left[1-\frac{2E^{2}}{b^{4}(r)\left(f(r)-1\right)}\right]^{-\frac{1}{2}}\\,
\end{equation}
where $r_b$ is the location of the boundary and we have use the relation $r^{'} = dr/dy$. In the small $E$ limit, we expand eq.(19) up to the leading order and rewrite the string action as,
\begin{equation}
S \simeq  \frac{L^-}{\sqrt{2} \pi \alpha^{'}} \int^{r_b}_{r_{min}}  dr
 b^{2}(r)\sqrt{\frac{f(r)-1}{f(r)}}\left[1+\frac{2E^{2}}{b^{4}(r)\left(f(r)-1\right)}\right]\\.
\end{equation}
This action is divergent and to eliminate the divergence it should be subtracted by the inertial mass of
two free quarks given by,
\begin{equation}
S_0 =  \frac{L^-}{\sqrt{2} \pi \alpha^{'}}
\int^{r_b}_{r_{min}} dr b^{2}(r)\sqrt{\frac{f(r)-1}{f(r)}}\,.
\end{equation}
Here, we have used the gauge $ x^- = \tau$ and $ r = \sigma$.
On the other hand, the distance between two quarks is obtained by integrating of eq.(17) as,
\begin{equation}
\frac{L}{2} = \int^{r_b}_{r_{min}} dr
\frac{\sqrt{2}E}{\sqrt{b^{4}(r)f(r)\left(f(r)-1\right)}}\left[1-\frac{2E^{2}}{b^{4}(r)\left(f(r)-1\right)}\right]^{-\frac{1}{2}}.
\end{equation}
Again by expanding this equation in terms of $E$ and considering the small $L$ limit, we obtain the following relation,
\begin{equation}
\frac{L}{2E} \simeq \int^{r_b}_{r_{min}}dr \sqrt{\frac{2}{b^{4}(r)f(r)\left(f(r)-1\right)}}\,.
\end{equation}
Also we can expand $E$ up to the leading order of $L$ to reach,
\begin{equation}
E\simeq \frac{L}{\sqrt{2} \int^{r_b}_{r_{min}}
\frac{dr}{b^{4}(r)f(r)\left(f(r)-1\right)}}\,.
\end{equation}
The the regularized string worldsheet action is then given by,
\begin{equation}
 S_I = S - S_0 = \frac{L^-L^2}{8\sqrt{2} \pi \alpha^{'}}\frac{1}{\int^{r_b}_{r_{min}}
  \frac{dr}{b^{2}(r)\sqrt{f(r)(f(r)-1)}}}\,,
\end{equation}
where we have used eqs.(23) and (24).
Finally, we obtain the the following equation for the jet quenching parameter for the scaling metric with hyperscaling violation,
\begin{equation}
\hat{q}_{LV} = \frac{1}{\pi \alpha^{'}}\left({\int^{r_b}_{r_{min}}
\frac{dr}{b^{2}(r)\sqrt{f(r)(f(r)-1)}}}\right)^{-1}\,.
\end{equation}
In the conformal limit ($\theta = 0$ and $z = 1$) this equation reduces to,
\begin{equation}
\hat{q}_{conf.} = \sqrt{2\lambda}\pi^{\frac{3}{2}} T^{3} \frac{\Gamma[\frac{3}{4}]}{\Gamma[\frac{5}{4}]}\,,
\end{equation}
which gives $\hat{q}_{conf.} =1.95 GeV^2/fm$ for $T = T_c$ and $\lambda=5.5$.
The Hawking temperature associated to the black hole metric (9) is given by,
\begin{equation}
T = \frac{1}{4\pi}\left|\frac{g^{'}_{00}}{\sqrt{-g_{00}g_{11}}}\right|_{r = r_h} = \frac{f_{0}}{8\pi l}(\frac{r_h}{l})^{\frac{z}{z-2}}\left|\frac{2z -3\theta+6}{z-2}\right|\,.
\end{equation}
In order to evaluate the jet quenching parameter of eq.(26) we should first determine the allowed region for parameters $z$ and $\theta$.
In this paper, we consider the space boundary at $r=\infty$, so the parameter $\alpha$ is negative. In this case, to have an
acceptable geometry satisfying the Gubser conditions of eq.(7), the parameter $\beta$ should be positive.\\
Also, to avoid divergence of the integrand in eq.(26) at $r\rightarrow \infty$ and to have a well-defined behavior for the jet quenching parameter,
we choose $\frac{5}{6}\leq\beta<1$ and $-3\leq\gamma\leq-2$.
These constraints together imply that the allowed geometries have the following $z$ and $\theta$ range,
\begin{eqnarray}
-\frac{3}{2}\theta+7\leq &z&\leq-4  \,\,\,\,\,\,\,\,\,\,\,\,\,\,\,\,\,\,\,\,\,\,\,\,\,\,\,\,for\,\,\,\,\,\,\,\,\,  \frac{22}{3}\leq\theta\leq\frac{34}{3}\,, \nonumber\\
-\frac{3}{2}\theta+7\leq &z&\leq -\frac{3}{4}\theta + \frac{9}{2} \,\,\,\,\,\,\,\,\,\,\,\,for\,\,\,\,\,\,\,\,\, \theta>\frac{34}{3}\,.
\end{eqnarray}
The allowed geometries with above constraints are shown in the $(z,\theta)$ plane in fig.1.\\
\begin{figure}
\centerline{\includegraphics[width=12cm]{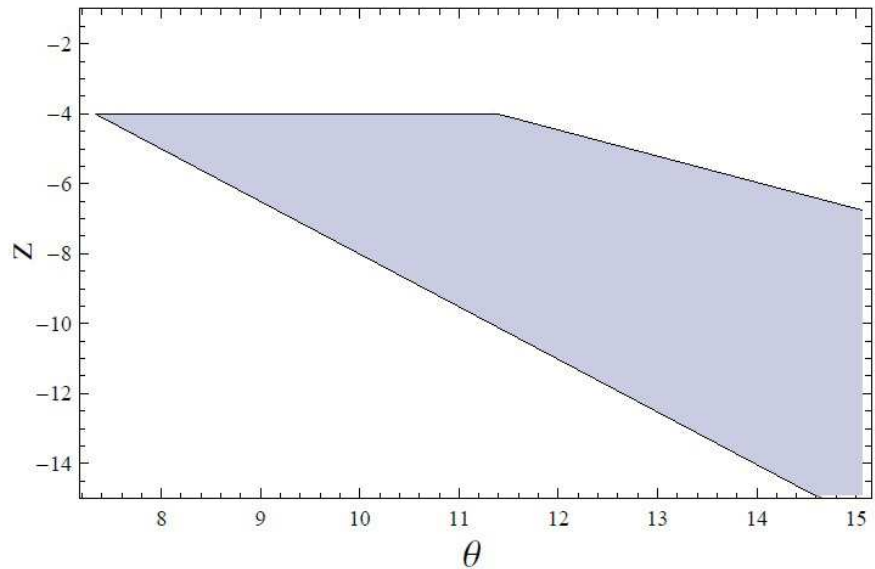}}
\caption{generalized Lifshitz geometries in the $(z,\theta)$ plane with $\alpha<0$\,, $\frac{5}{6}\leq\beta<1$ and $-3\leq\gamma\leq-2$.}
\end{figure}
To compare our results with those from conformal case and experimental data and study the behavior of the jet quenching parameter $\hat{q}_{LV}$, we plot it numerically for $z=-8$ and $\theta=10$ case and $q_{conf.}$ as a function of temperature in Fig.2.\\
\begin{figure}
\centerline{\includegraphics[width=12cm]{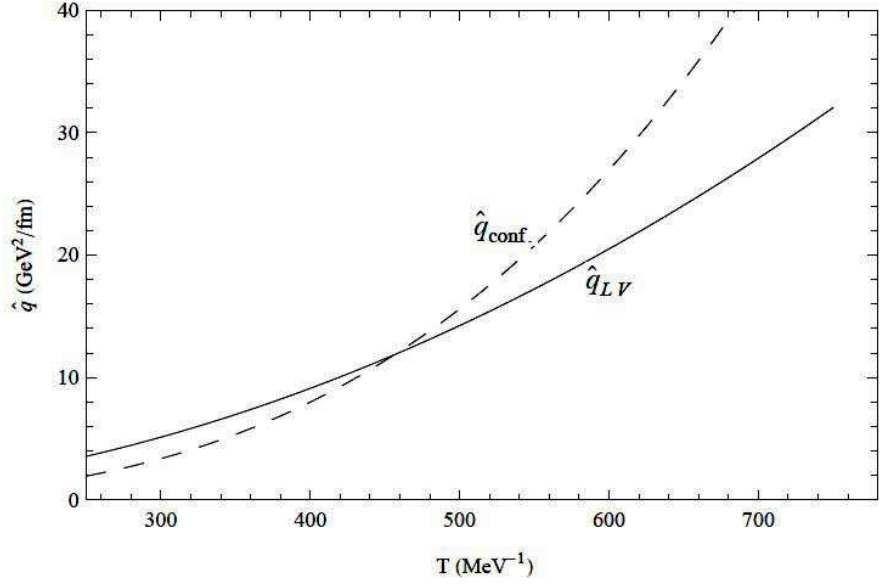}}
\caption{jet quenching parameter as a function of T for the scaling metric with hyperscaling violation (solid line) and
the conformal case (dashed line) for $z=-8$ and $\theta=10$ case.}
\end{figure}
From Fig.2, one can see that at $T_c \simeq 250\,MeV$,  $\hat{q}_{LV}\simeq 3.56\,GeV^2/fm$ which is larger than $\hat{q}_{conf.}$. As temperature increases,
$\hat{q}_{LV}$ decreases and at $T \simeq 460\, MeV$ it equals to $\hat{q}_{conf.}$ and then it gets smaller than $\hat{q}_{conf.}$.\\
\begin{figure}
\centerline{\includegraphics[width=12cm]{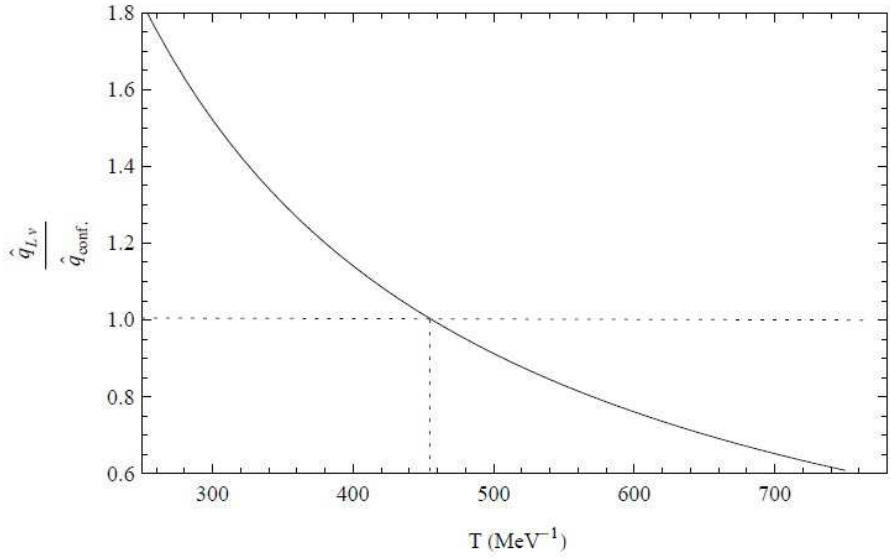}}
\caption{$\hat{q}_{LV}/\hat{q}_{conf.}$ as a function of temperature for $z=-8$ and $\theta=10$ case.
For $T<T_0$ the ratio is larger than 1 for $T>T_0$, the ratio is less than 1.}
\end{figure}
In Fig.3 we demonstrate $\hat{q}_{LV}/\hat{q}_{conf.}$ as a function of temperature for $z=-8$ and $\theta=10$ case. The numerical values of $\hat{q}_{LV}$ for different temperatures
are shown in Table.1.\\
\begin{table}
\caption{Jet quenching parameter $\hat{q}_{LV}$ in the units of $GeV^2/fm$ for different values of temperature. The temperature is given in the units of critical temperature $T_c$.}
\vspace{0.3in}
\centering
\begin{tabular}{|c|c|c|c|c|c|}
\hline
&$T=T_{c}$ &$T=1.5T_{c}$ & $T=2T_{c}$ & $T=2.5T_{c}$ &$T=3 T_{c}$ \\

\hline

$\hat{q}_{LV}$&  3.56  & 8.01  &  14.24  & 22.25 & 32.05\\
\hline
\end{tabular}
\end{table}
From the numerical values of the jet quenching parameter, it could be realized that in the range of $300<T<500$, our results for $\hat{q}_{LV}$ is in consistent with those obtained
from RHIC with $\bar{\hat{q}}\simeq 5-15\,GeV^{2}/fm$ [59-60].\\
Now, we assume two different cases to examine the behavior of the jet quenching parameter $\hat{q}_{LV}$ as a function of $T$ for different values of parameters $z$ and $\theta$.\\
\begin{itemize}

\item To evaluate the jet quenching parameter as a function of temperature for different values of $\theta$ and fixed Lifshitz exponent ($z = -8$). This case is shown in Fig.4. From this plot, one can figure out that $\hat{q}_{LV}$ becomes large as $\theta$ increases.
\item To evaluate the jet quenching parameter as a function of temperature for different values of $z$ and fixed hyperscaling exponent ($\theta = 10$). This case is shown in Fig.5. In this case, $\hat{q}_{LV}$ decreases as $\theta$ increases.

\end{itemize}

\begin{figure}
\centerline{\includegraphics[width=12cm]{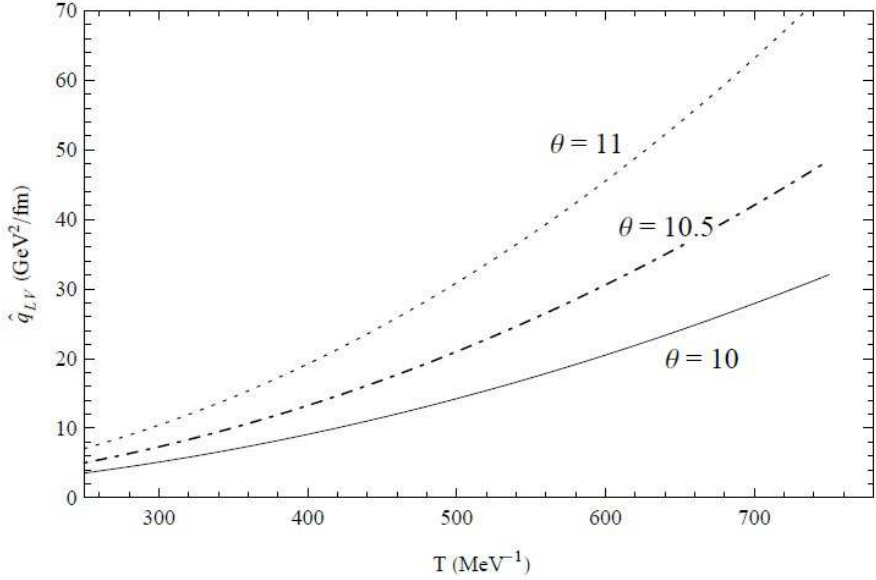}}
\caption{ $\hat{q}_{LV}$ as a function of temperature for different values of $\theta$ and constant value of Lifshitz exponent $z = -8$. As $\theta$ increases, the jet quenching parameter increases.}
\end{figure}

\begin{figure}
\centerline{\includegraphics[width=12cm]{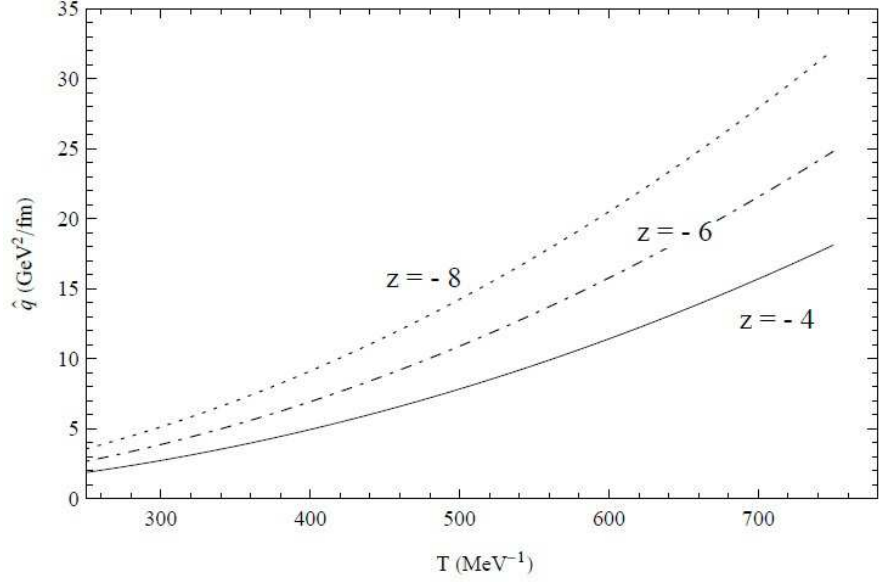}}
\caption{$\hat{q}_{LV}$ as a function of temperature for different values of $z$ and fixed hyperscaling exponent $\theta = 10$. As $z$ increases, the jet quenching parameter decreases.}
\end{figure}
Then we plot the jet quenching parameter as a function of hyperscaling exponent $\theta$ for different values of $T$ and fixed Lifshitz exponent $z = -8$ in Fig.6. From this figure, it is clear that $\hat{q}_{LV}$ is an increasing function of $\theta$.\\

\begin{figure}
\centerline{\includegraphics[width=12cm]{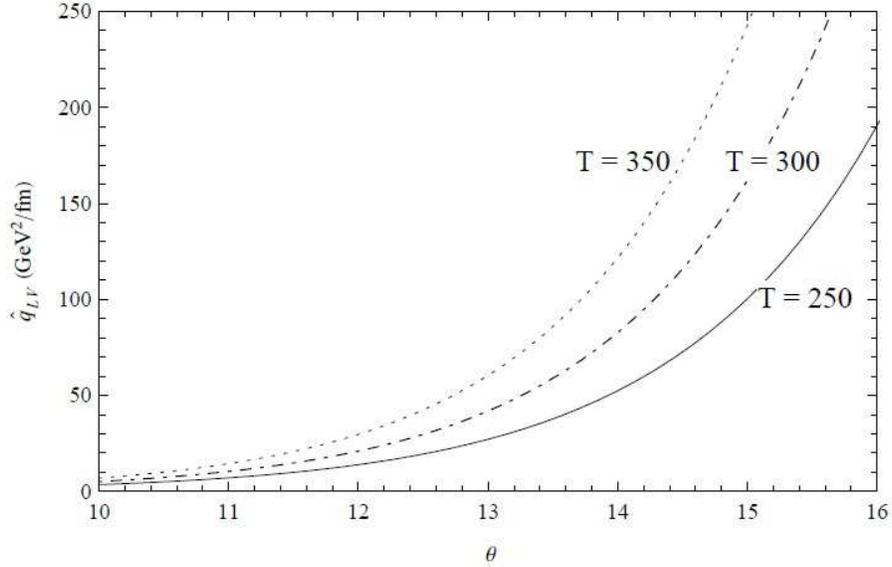}}
\caption{$\hat{q}_{LV}$ as a function of $\theta$ for different values of $T$ and fixed $z = -8$. As $\theta$ increases, the jet quenching parameter increases.}
\end{figure}

Finally $\hat{q}_{LV}$ as a function of Lifshitz exponent is depicted in Fig.7 for different values of $T$ and fixed hyperscaling violation exponent $\theta = 10$. Unlike the previous case, it is a decreasing function of $z$.\\
\begin{figure}
\centerline{\includegraphics[width=12cm]{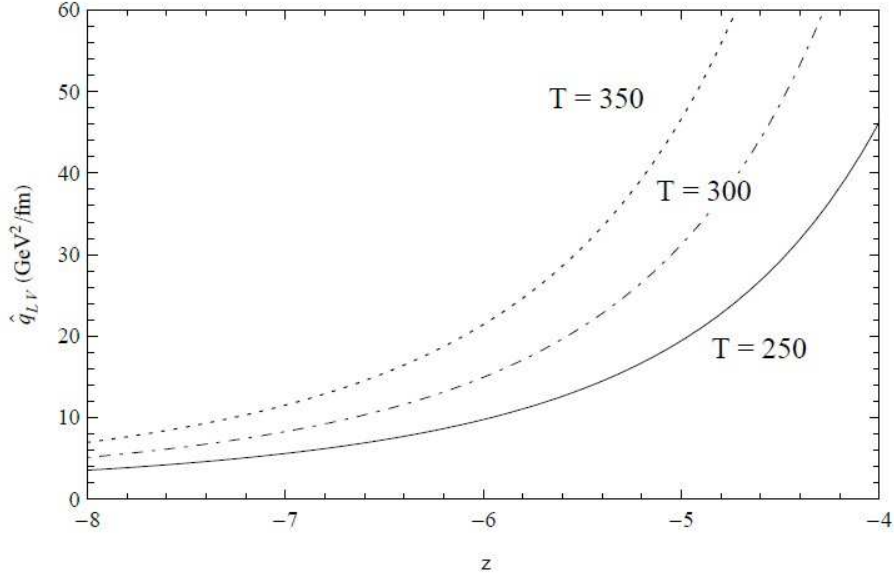}}
\caption{$\hat{q}_{LV}$ as a function of $z$ for different values of $T$ and fixed $\theta = 10$. As $z$ increases, the jet quenching parameter decreases.}
\end{figure}

\section{Effect of constant electric field}
In the previous section, we evaluated the jet quenching parameter in the scaling background with hyperscaling violation at finite temperature introduced in ref.[57]. Now, we proceed to study the effect of a constant electric field on the jet quenching parameter following the method proposed in ref.[15].
In this set up, the constant B-field is considered to be along the $x^1$ and $x^2$ direction. Due to the fact that
only the field strength is involved in equations of motion, this
ansatz could be a good solution to supergravity and this is the minimal setup for the dual field
theory to study the B-field correction. The constant B-field is added to the line element of eq.(9) by the following two form,
\begin{equation}\label{ads_BH}
B=B_{01}dt\wedge dx_1 + B_{12}dx_1\wedge dx_2,
\end{equation}
where $B_{01}={\cal E}$ and $B_{12}={\cal H}$ are constant NS-NS antisymmetric electric and magnetic fields.\\
In this paper we consider $B_{12}=0$ and study the effect of constant electric field on the evolution jet quenching parameter.
Similar to the previous section, we use the static gauge $\tau = x^-$, $\sigma
= x^2$ and the light-cone coordinates of eq.(12). At the presence of the electric field, the string action is described by,
\begin{eqnarray}
&&S=-\frac{1}{2\pi\alpha'}\int{d\tau
d\sigma\sqrt{-det(g+b)_{\alpha\beta}}},
\end{eqnarray}
where $\alpha,\beta=\tau,\sigma$ and $g_{\alpha\beta}$ is given in eq.(14). In the light-cone coordinates,
the induced b-field on the string worldsheet, $b_{\alpha\beta} = B_{\mu\nu}\partial_{\alpha}X^{\mu}\partial_{\beta}X^{\nu}$, is obtained as,
\begin{equation}
b_{\alpha\beta} =\left(\begin{array}{cc}\frac{{\cal E}}{2}& 0\\ 0 & 0\end{array}\right)\,.
\end{equation}
Putting eqs.(14) and (32) in eq.(31) for the string action leads to,
\begin{equation}
S =\frac{L^-}{\sqrt{2} \pi \alpha^{'}} \int^{\frac{L}{2}}_{0}  dy b^{2}(r)\sqrt{\left(f(r)-1-\frac{{\cal E}}{b^2(r)}\right)\left(1+\frac{r^{'2}}{f(r)}\right)}\,.
\end{equation}
Also, the equation of motion for $r$ is founded to be,
\begin{equation}
{r^{'2}} = f(r)\left[\frac{b^4(r)}{2E^2}\left(f(r)-1-\frac{{\cal E}}{b^2(r)}\right)-1\right]\,.
\end{equation}
Inserting this equation into eq.(33) and taking the $E\rightarrow 0$ limit, we obtain the following equation for the string action,
\begin{equation}
S =  \frac{L^-}{\sqrt{2} \pi \alpha^{'}} \int^{\infty}_{r_{min}}  dr
 b^{2}(r)\sqrt{\frac{f(r)-1-\frac{{\cal E}}{b^2(r)}}{f(r)}}\left[1+\frac{2E^{2}}{b^{4}(r)\left(f(r)-1-\frac{{\cal E}}{b^2(r)}\right)}\right]\,,
\end{equation}
and the following equation for the self energy of two quarks,
\begin{equation}
S_0 =  \frac{L^-}{\sqrt{2} \pi \alpha^{'}}
\int^{\infty}_{r_{min}} dr b^{2}(r)\,f^{-1}(r)\sqrt{f(r)-1-\frac{{\cal E}}{b^2(r)}}\,.
\end{equation}
Also the distance between two end points of string becomes,
\begin{equation}
L = 2\sqrt{2}E\int^{\infty}_{r_{min}}dr \left[b^{4}(r)f(r)\left(f(r)-1-\frac{{\cal E}}{b^2(r)}\right)\right]^{-\frac{1}{2}}\,.
\end{equation}
Finally, we obtain the jet quenching parameter as,
\begin{equation}
\hat{q}_{LV} = \frac{1}{\pi \alpha^{'}}\left({\int^{\infty}_{r_{min}}
\frac{dr}{b^{2}(r)\sqrt{f(r)(f(r)-1-\frac{{\cal E}}{b^2(r)})}}}\right)^{-1}\,,
\end{equation}
where $r_{min}$ is obtained numerically as the solution of the following equation,
\begin{equation}
f(r_{min})-1-\frac{{\cal E}}{b^2(r_{min})} = 0\,,
\end{equation}
Now we proceed to evaluate the jet quenching parameter numerically and study its behavior at the presence of a constant electric field.
For this purpose, we consider two different cases with numerical values $z = -8$ for Lifshtz exponent and $\theta = 10$ for hyperscaling violation exponent.\\
First, we numerically evaluate the jet quenching parameter as a function of temperature for different values of ${\cal E}$. From this plot, one can figure out that $\hat{q}_{LV}$ becomes large as $T$ increases. This case is shown in Fig.8. \\
\begin{figure}
\centerline{\includegraphics[width=12cm]{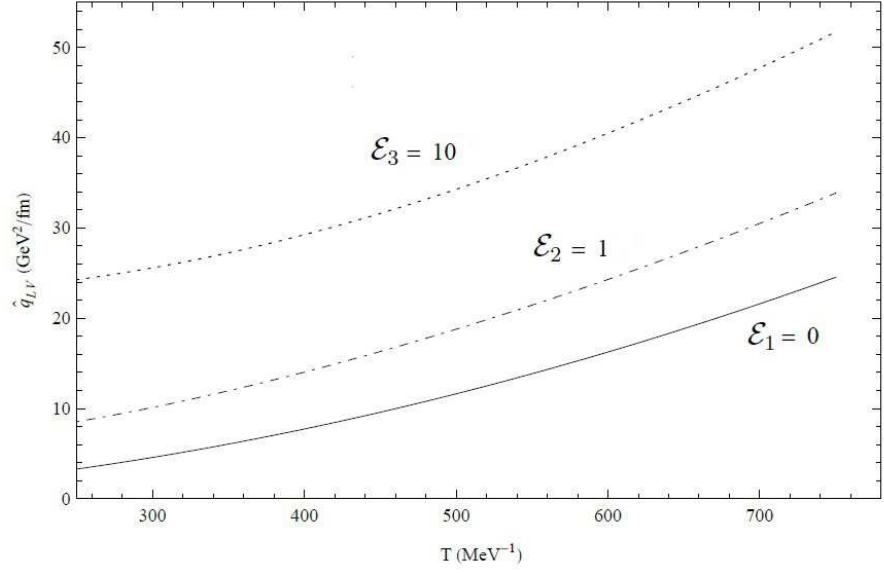}}
\caption{$\hat{q}_{LV}$ as a function of $T$ for different values of ${\cal E}$ (${\cal E}_1 < {\cal E}_2 < {\cal E}_3$). As $T$ increases, the jet quenching parameter increases.}
\end{figure}
Second, we numerically evaluate the jet quenching parameter as a function of ${\cal E}$ for different values of $T$ in the units of $T_c$. From this plot we can see that in constant temperature, $\hat{q}_{LV}$ increases as $T$ increases. This case is shown in Fig.9.
\begin{figure}
\centerline{\includegraphics[width=12cm]{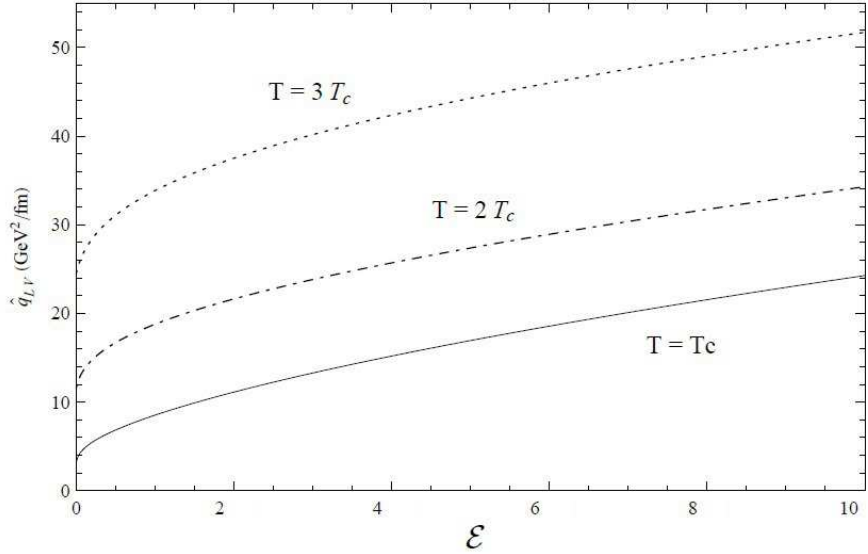}}
\caption{$\hat{q}_{LV}$ as a function of ${\cal E}$ for different values of $T$ in the units of $T_c$. The jet quenching parameter increases, as ${\cal E}$ increases.}
\end{figure}

\section{Results and discussion}
Jet quenching of partons produced at RHIC with high transverse momentum is one of the interesting properties of the strongly-coupled plasma. It is possible to determine this quantity
using the gauge/gravity duality for  gauge theories at finite temperature. In this paper we analyzed the behavior of jet quenching parameter in the background metric which is
covariant under a generalized Lifshitz scaling symmetry and has a generic Lorentz violating form at finite temperature.\\
In section 3, we used the holographic description to study the jet quenching parameter at finite temperature. For this purpose, we determined the appropriate range for $z$ and $\theta$ and evaluate the jet quenching parameter numerically in terms of $T$, $z$ and $\theta$ parameters. We considered the space boundary at $r=\infty$, so the negative $\alpha$ parameter.
In this case, to have an acceptable geometry satisfying the Gubser conditions, the parameter $\beta$ should be positive. To avoid divergence of the jet quenching integrand at $r\rightarrow \infty$ and to have a well-defined behavior for the jet quenching parameter, we chose $\frac{5}{6}\leq\beta<1$ and $-3\leq\gamma\leq-2$. We ploted $\hat{q}_{LV}-T$ numerically for $z=-8$ and $\theta=10$ as a function of temperature in Fig.2, which is qualitatively similar to references [18] and [20]. At $T_c \simeq 250\,MeV$,  $\hat{q}_{LV}\simeq 3.56\,GeV^2/fm$ which is larger than $\hat{q}_{conf.}$. As temperature increases,
$\hat{q}_{LV}$ decreases and at $T \simeq 460\, MeV$ it equals to $\hat{q}_{conf.}$ and then it gets smaller than $\hat{q}_{conf.}$. From the numerical values of the jet quenching parameter, we found that in the range of $300<T<500$, our results for $\hat{q}_{LV}$ is in consistent with results from RHIC, $\bar{\hat{q}}\simeq 5-15\,GeV^{2}/fm$ [59-60]. We evaluated the jet quenching parameter as a function of temperature for different values of $\theta$ and $z$ in Fig.4. From this plot, one can figure out that $\hat{q}_{LV}$ becomes large as $\theta$ increases.
Then for different values of $z$ and constant value $\theta$ we plotted the jet quenching parameter as a function of temperature (Fig.5). In this case, $\hat{q}_{LV}$ decreases as $\theta$ increases. Then we plotted the jet quenching parameter as a function of hyperscaling exponent $\theta$ for different values of $T$ and fixed Lifshitz exponent $z$ in Fig.6. From this figure, we found that $\hat{q}_{LV}$ is an increasing function of $\theta$. Also $\hat{q}_{LV}$ as a function of $z$ is depicted in Fig.7 for different values of $T$ and constant value of $\theta$. Unlike the previous case, it is a decreasing function of $z$.\\\\
Finally, we added a constant B-field (electric filed) to the background metric and estimate the jet quenching parameter dependence on the Hawking temperature and also the electric field for $z = -8$ and $\theta = 10$. From fig.8, one can figure out that $\hat{q}_{LV}$ becomes large as $T$ increases. Then, from fig.9 we can see that in constant temperature, $\hat{q}_{LV}$ is an increasing function of ${\cal E}$.


\end{document}